\documentclass[aps,twocolumn]{revtex4}
\usepackage{amsmath}
\usepackage{amsfonts}
\usepackage{mathrsfs}
\usepackage{graphicx}
\usepackage{times}
\usepackage{color}
\usepackage{multirow}
\begin{document}

\title{A toolbox for elementary fermions with a dipolar Fermi gas in a $3$D optical lattice}

\author{Shuai Li}
\affiliation{MOE Key Laboratory for Nonequilibrium Synthesis and Modulation of Condensed Matter,Shaanxi Province Key Laboratory of Quantum Information and Quantum Optoelectronic Devices, School of Physics, Xi'an Jiaotong University, Xi'an 710049, China}

\author{Maksims Arzamasovs}
\affiliation{MOE Key Laboratory for Nonequilibrium Synthesis and Modulation of Condensed Matter,Shaanxi Province Key Laboratory of Quantum Information and Quantum Optoelectronic Devices, School of Physics, Xi'an Jiaotong University, Xi'an 710049, China}

\author{Hongrong Li}
\affiliation{MOE Key Laboratory for Nonequilibrium Synthesis and Modulation of Condensed Matter,Shaanxi Province Key Laboratory of Quantum Information and Quantum Optoelectronic Devices, School of Physics, Xi'an Jiaotong University, Xi'an 710049, China}

\author{Fuli Li}
\affiliation{MOE Key Laboratory for Nonequilibrium Synthesis and Modulation of Condensed Matter,Shaanxi Province Key Laboratory of Quantum Information and Quantum Optoelectronic Devices, School of Physics, Xi'an Jiaotong University, Xi'an 710049, China}

\author{Bo Liu}
\email{liubophy@gmail.com}
\affiliation{MOE Key Laboratory for Nonequilibrium Synthesis and Modulation of Condensed Matter,Shaanxi Province Key Laboratory of Quantum Information and Quantum Optoelectronic Devices, School of Physics, Xi'an Jiaotong University, Xi'an 710049, China}

\begin{abstract}
There has been growing interest in investigating properties of elementary particles predicted by the standard model. Examples of such studies include exploring their low-energy analogs in condensed matter system, where they arise as collective states or quasiparticles. Here we show that a toolbox for systematically engineering the emergent elementary fermions, i.e., Dirac, Weyl and Majorana fermions, can be built in a single atomic system composed of a spinless magnetic dipolar Fermi gas in a $3$D optical lattice. The designed direction-dependent dipole-dipole interaction leads to both the basic building block, i.e, in-plane $p+ip$ superfluid pairing instability
and the manipulating tool, i.e, out-of-plane Peierls instability. It is shown that the Peierls instability provides a natural way of tuning the topological nature of $p+ip$ superfluids and thus transform the fermion's nature between distinct emergent particles. Our scheme should open up a new thrust towards searching for elementary particles through manipulating the topology.
\end{abstract}

\maketitle

Fundamental particles are either the building blocks of matter, called fermions, or the mediators of interactions, called bosons. These elementary particles can be understood within the framework of the relativistic quantum field theory~\cite{2010_book,2005_book}, such as Dirac, Weyl and Majorana fermions~\cite{Dirac_1,Dirac_2,Klein,Weyl,Majorana}. However, only Dirac fermions have been observed as elementary particles in nature so far. For many years now, another promising approach to observe particle properties that have no realization in elementary particles is the investigation of their low-energy analogous quasiparticles, such as in condensed matter or atomic systems. It paves a new way for exploring fundamental particles without paying the steep price of a high-energy particle collider and thus has attracted tremendous research interests in various fields of physics. There have been several exciting progresses on searching the emergent Dirac, Weyl and Majorana fermions. Recent examples include graphene~\cite{Geim_RevModPhys,Goerbig_RevModPhys,Neto_RevModPhys,Firsov_Nature_2005,
Philip_Nature_2005,Geim_NatPhys_2006}, several topological phases in solids containing quantum Hall states, topological insulator/superconductor and etc.~\cite{2003_Volovik_book,RevModPhys1,RevModPhys2}. Another exciting perspective is the recent realization of artificial materials, such as cold atoms~\cite{GalitskiiSpielman,2011_RevModPhys,LinJimenezGarciaSpielman,WuZhangSunXuWangJiDengChenLiuPan,2012_Cheuk_PhysRevLett, 2012_Wang_PhysRevLett,2014_Jotzu_Nature,DucaLiReitterBlochEtal,2013_Aidelsburger_PhysRevLett, 2013_Miyake_PhysRevLett,2013_cheng_Natphys,2012_huizhai_IJMP,2014_Qizhou_PhysRevA,2014_huizhai_PhysRevA,2014_cooper_PhysRevA,
2007_Bloch_PRL,2011_Hemmerich_NatPhys,2012_Sengstock_Natphys,2015_Hemmerich_PhysRevLett}, or photonic crystals~\cite{2013_Nature,2013_NatureMat}. However, finding a single material that can systematically transform the fermion's nature and realize distinct elementary fermions in its equilibrium state is highly non-trivial and still stands as an obstacle to be yet overcome.

Here we show that ultracold gases of magnetic dipolar atoms or polar molecules, as presently developed in the laboratory, provide us new opportunities for constructing a toolbox for systematically engineering all three kinds of elementary fermions listed above. The attractiveness of this idea rests on the fact that the strength and even the sign of dipolar interaction in cold atoms are highly tunable~\cite{2012_Baranov_Reviews}. The direction of dipole moments can be fixed by applying an external magnetic field. Let the external field be orientated at a small angle with respect to the $xy$-plane and rotate fast around the $z$-axis. The time-averaged interaction between dipoles is isotropically attractive in the $xy$-plane and repulsive along the $z$-direction. Such a scheme has been realized in the experimental system of dysprosium
atoms~\cite{Tang_2018_PhysRevLett}. In general, the $xy$-plane attraction is expected to cause superfluid pairing instability and leads to a in-plane $p+ip$ superfluid in a spinless dipolar Fermi gas. The repulsion should restrict the pairing along $z$-direction and results in the Peierls instability in the presence of lattice potential. Such a spontaneously formed density modulation provides a natural tool to manipulate the topological nature of $p+ip$ superfluids and thus allows us to build a toolbox for systematically engineering all the above three kinds of elementary fermions through tuning the topology of our proposed single atomic system. Such a heuristically argued result is indeed confirmed by our detailed analysis through the model to be introduced below.

\begin{figure*}
  \begin{center}
  \includegraphics[width=1\textwidth]{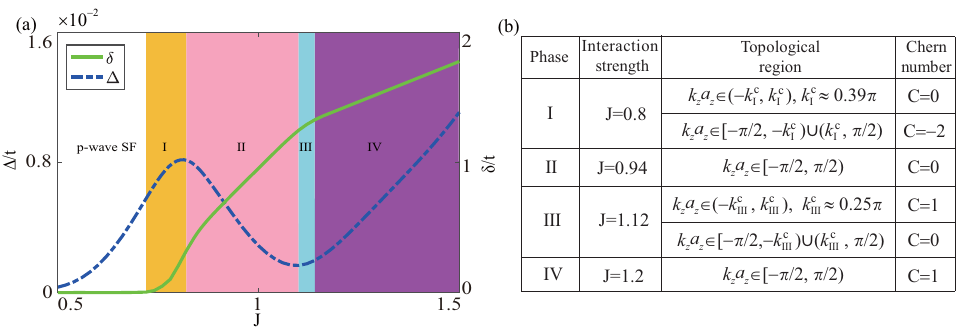}
  \end{center}
  \caption{(a) Zero-temperature phase diagram as a function of dipolar interaction
   strength at average filling $n_0=0.6$. The dashed and solid lines stand for the pairing and CDW order parameters, respectively. The spontaneously formed density modulation serves as a natural tool to manipulate the topological nature of the
   system, leading to distinct topological phases marked by various colors.
   (b) A table of examples showing the distinct topological nature of various phases in (a). Other parameters are $t_z/t=0.5$, $a_z/a=3.5$.}\label{fig:phase}
   \end{figure*}

{\textbf{{Effective model}}

Consider a spinless dipolar Fermi gas, such as $^{161}Dy$ ~\cite{2012_Luming_PRL,2014_Benjamin_PRA} or $^{167}Er$ ~\cite{2014_Grimm_PRL,2014_Ferlaino_arxiv}, subjected to an external
rotating magnetic field $\mathbf{B}(t)=B[{\hat{z}}\cos {\varphi }+\sin {\varphi }({\hat{x}}\cos {\Omega }t+{\hat{y}}\sin {\Omega }t)]$, where $\Omega$ is the rotation frequency, $B$ is the magnitude of magnetic field, the rotation axis is $z$. $\varphi $ is the angle between the magnetic field and $z$-axis. In strong magnetic fields, dipoles are aligned parallel to $\mathbf{B}(t)$. With fast rotations, i.e., $\Omega$ being much larger than typical frequencies of particle motion, and simultaneously much smaller than level splitting in the field, the effective interaction between dipoles is the time-averaged interaction
$V(\mathbf{r})={\frac{d^{2}(3\cos ^{2}\varphi -1)}{2r^{3}}(1-3\cos
^{2}\theta )}\equiv \frac{d^{\prime 2}}{r^{3}}{(1-3\cos ^{2}\theta )},$
where $d^{\prime 2}\equiv d^{2}{\frac{(3\cos ^{2}\varphi -1)}{2}}$ with the
magnetic dipole moment $d$. {$\mathbf{r}$ is the vector connecting two
dipolar particles, and $\theta $ is the angle between $\mathbf{r}$ and the $z$-axis.}
The effective in-plane attraction is created by making
$\cos \varphi <\sqrt{1/3}$, which can be realized by changing the amplitudes of static and rotating parts of magnetic field. We further consider these dipolar atoms loaded in a three dimensional optical lattice $V_{opt}(\mathbf{r})=-V_{0}[\cos ^{2}(k_{Lx}x)+\cos ^{2}(k_{Ly}y)]-V_{0z}\cos
^{2}(k_{Lz}z)$, where $k_{Lx},k_{Ly}$ and $k_{Lz}$ are wave vectors of
laser fields and the corresponding lattice constants are defined as $%
a_{x}=\pi /k_{Lx},$ $a_{y}=\pi /k_{Ly}$ and $a_{z}=\pi /k_{Lz}$. Here $V_{0}$ and $V_{0z}$ are lattice depths in the $xy$-plane and $z$-direction, respectively. In this work, we consider an anisotropic 3D lattice with $a_{x}=a_{y}\equiv a <a_{z}$. When the lattice depths are large enough, the system can be described by the following Fermi-Hubbard model in the tight-binding regime
\begin{eqnarray}
\textstyle
{\bf H} &=& -\sum_{\alpha =x,y,z}\sum_{i}t_{\alpha }(c_{i}^{\dag }c_{i+{e}_{\alpha
}}+h.c.)-\mu \sum_{i}c_{i}^{\dag }c_{i}  \notag \\
&+& \frac{1}{2}\sum_{i\neq j}V_{i-j}c_{i}^{\dag }c_{j}^{\dag }c_{j}c_{i},
\label{Haml}
\end{eqnarray}
where $t_{x}=t_{y}\equiv t$ and $t_{z}$ are the hopping amplitude describing
tunneling along $x,y$ and $z$ directions, respectively. $i\equiv(i_x,i_y,i_z)$ is the site index denoting the lattice site $\mathbf{R}%
_{i}\equiv(ai_x,ai_y,a{_z}i_z)$. $\mu $ is the chemical potential and ${e}_{\alpha
}$ represents the unit
vector. The dipole-dipole interaction is given by $V_{i-j}=d^{\prime 2}\frac{|\mathbf{R}_{i}-\mathbf{R}%
_{j}|^{2}-3(i_{z}-j_{z})^{2}a_z^{2}}{|\mathbf{R}_{i}-\mathbf{R}_{j}|^{5}}$.

To study the many-body instabilities of the Hamiltonian Eq.~\eqref{Haml}, we have  performed  both mean-field theory and Monte Carlo (MC) simulations. By employing the self-consistent Hartree-Fock-Bogoliubov (HFB) method, the zero-temperature phase diagram is obtained as shown in Fig.~\ref{fig:phase}(a).  Furthermore, distinct many-body phases in Fig.~\ref{fig:phase}(a) were identified through MC simulations. In the HFB method, the Peierls instability can be described by writing the density distribution of the system as $n_{i}=n_{0}+%
C\cos (\mathbf{Q}\cdot \mathbf{R}_{i})$, where $\mathbf{Q}$
represents the periodicity of density pattern and  $n_{0}=\sum%
\limits_{i}\langle c_{i}^{\dagger }c_{i}\rangle / N_L$ is the average
filling with $N_L$ denoting total lattice site.
The order parameter describing this charge density wave (CDW) can thus be defined as $\delta _{\pm \mathbf{Q}}=V(\pm \mathbf{Q})C/2$, where $V(\mathbf{k})=\sum_{n\neq 0}V_{n}\exp (-i\mathbf{k}\cdot \mathbf{r}_{n})$.  We also introduce the superfluid pairing order parameter as $\Delta ({\mathbf{k}})=\frac{1}{N_L}\sum_{\mathbf{k'}}
V(\mathbf{k}-\mathbf{k'})<c_{-\mathbf{k'}}c_{\mathbf{k'}}>$
and $<...>$ means the expectation
value in the ground state.

Through minimizing the mean-field thermodynamic potential, order parameters defined above can be obtained (see details in Supplementary Materials (SM)).
We find that the pairing order parameter $\Delta({\bf k})$ behaves like in-plane $p+ip$ superfluids and can be described as
$\Delta({\bf k})= \Delta (\sin (k_xa)+i\sin (k_ya))$.
At the same time, we also find that
the CDW order is highly tunable
through simply varying the average filling. For instance, there is a region of $n_{0}$, i.e., $0.5\leq n_{0}<n_{A}\approx 0.67$, where $\mathbf{Q}$ is located at $(0,0,\pi /a_{z})$, indicating the period of $z$-directional
density modulation being $2a_{z}$. When further increasing $n_{0}$, i.e.,$n_{A} <n_{0}<n_{B}\approx 0.85$, the period of $z$-directional CDW order can be changed to $3a_{z}$,  characterized by $\mathbf{Q}$ fixed at $(0,0,{2\pi}/3a_{z})$. To further verify the existence of CDW and superfluid orders, we have performed a variational Monte Carlo (VMC)~\cite{MC1_PhysRevB,MC2_GROS,Pfaffian_PhysRevB,Jastrow_PhysRev,SR_PhysRevB} calculation on an $8 \times 8 \times 8$ lattice system with periodic boundary condition. Their existence can be captured by the long-ranged saturation behavior of the in-plane superfluid pairing correlation $P(\mathbf{R}_{\parallel})$ and the peak of structure factor $S(\mathbf{Q})$, respectively (see details in SM). It is shown that the results of the VMC simulation are consistent with the mean-field calculation, such as shown in Fig.~\ref{fig:phase}(a).

\begin{figure}[t]
\begin{center}
\includegraphics[width=8.5cm]{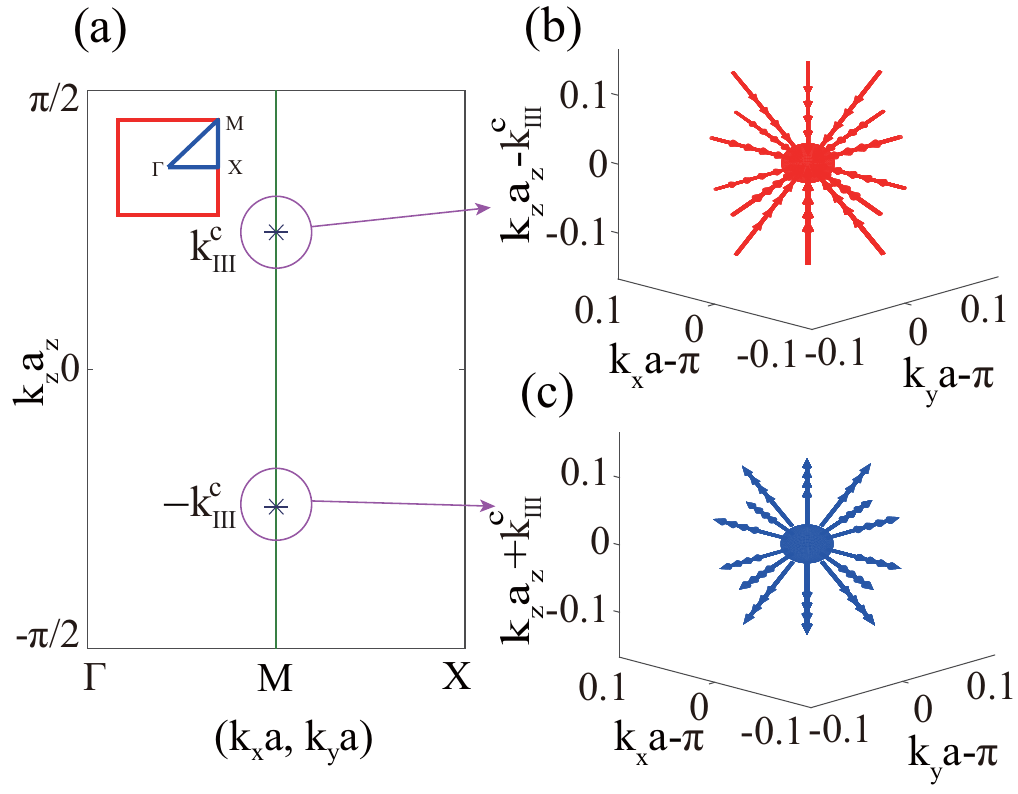}
\end{center}
\caption{ (a) Gapless points of the quasiparticle excitation in phase III. A contour in $(k_x,k_y)$-plane has been chosen as shown in the inset. (b)(c) Hedgehog-like topological defects formed by vector $\mathbf{d}$ around two Weyl nodes. The interaction strength $J=1.12$. Other parameters are the same as in Fig.~\ref{fig:phase}. } \label{fig:Weyl}
\end{figure}

\begin{figure}[t]
\begin{center}
\includegraphics[width=8.8cm]{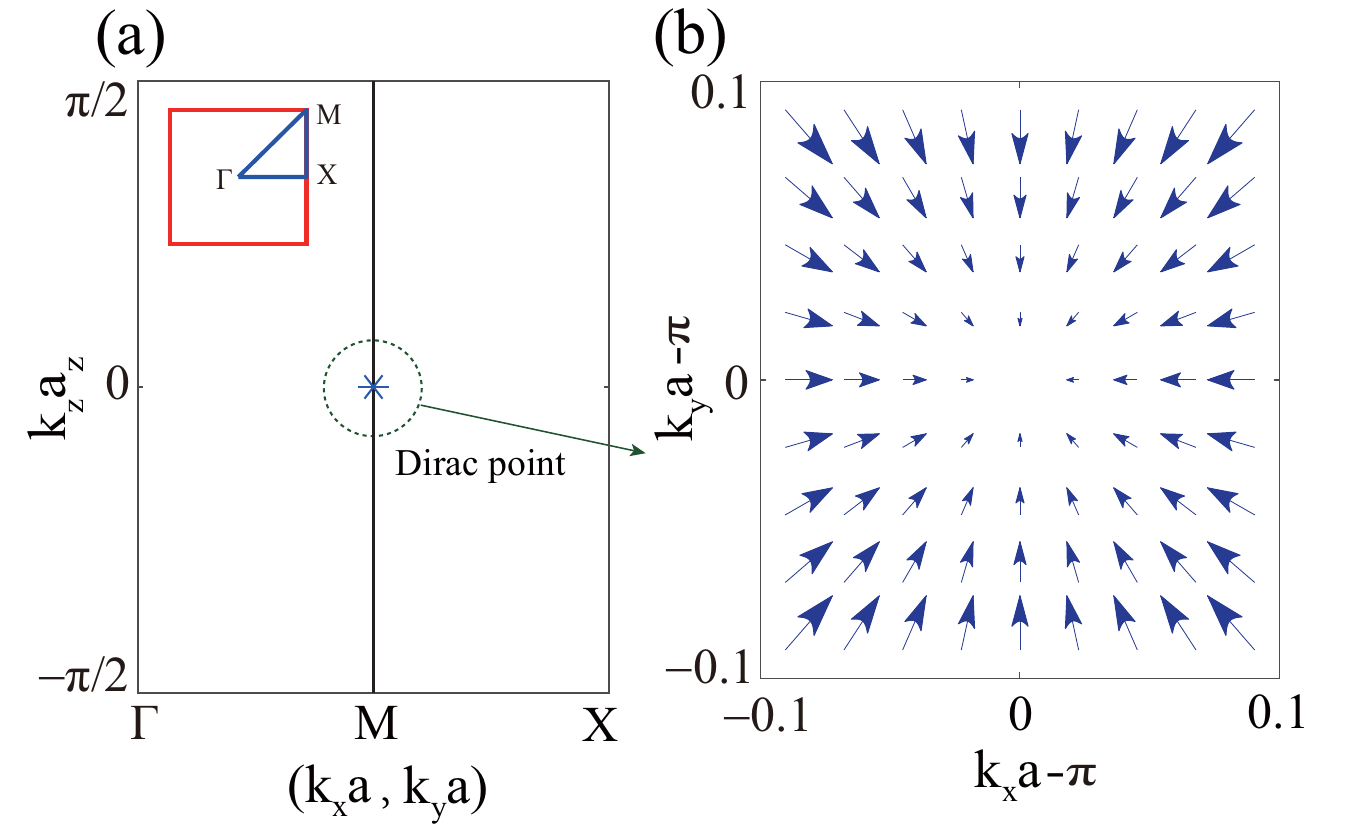}
\end{center}
\caption{ (a) Gapless point of the quasiparticle excitation on the
phase boundary between III and IV in Fig.~\ref{fig:phase}(a).
(b) Dirac topological defect formed by the vector $\mathbf{h}$ around the gapless point.  Other parameters are the same as in Fig.~\ref{fig:phase}. } \label{fig:Dirac}
\end{figure}

\textbf{Physical mechanism for tuning the topological nature}

To understand how we can utilize the spontaneously formed density modulation as a tool to manipulate the topological nature of the system, let us start with our basic building
block, i.e., in-plane $p+ip$ superfluids. It is known that the topology of $2$D $p+ip$ superfuilds can be changed by tuning the system filling and thus result in distinct topological regions: topological trivial and two distinct topological
regions with opposite chirality~\cite{2013_book}. In our proposed system, the spontaneously formed $z$-directional density modulation can serve as a natural tool to tune the fillings of $p+ip$ superfluid layers through effectively altering their respective chemical potential and thus changes their topological nature, which is  confirmed by our detailed analysis below.

For instance, let us consider the case of the
period of $z$-directional density modulation
being $2a_z$, i.e, $\mathbf{Q} =(0,0,\pi/a_z)$.
The topology of the system can be
understood through the Bogliubov-de Gennes (BdG) Hamiltonian. After applying a series of unitary transformations (see SM for details), the BdG Hamiltonian can be expressed in a clearer way as
\begin{eqnarray}
\mathscr{H}^{\pi}_{BdG}
&=& \left( \begin{array}
[c]{cccc}%
{E}^{\pi}_{1} & \Delta({\bf k}) & {0} & 0\\
\Delta^{\ast}({\bf k}) & -{E}^{\pi}_{1} & {0} & 0\\
{0} & {0} & {E}^{\pi}_{2} & \Delta({\bf k})\\
0 & 0 & \Delta^{\ast}({\bf k}) & -{E}^{\pi}_{2}%
\end{array}
\right) \nonumber \\
&\equiv &\left(
\begin{array}
[c]{cc}%
H'^{\pi}_{p-wave} & 0_{2\times2}\\
0_{2\times2} & H''^{\pi}_{p-wave}%
\end{array}
\right),
\end{eqnarray}
where $E^{\pi}_{1}({\bf k})=\frac{{\xi_{\mathbf{k}}+\xi_{\mathbf{k}+\mathbf{Q}}}}{2}%
-\sqrt{4{\delta}^{2}+(\frac{{\xi_{\mathbf{k}}-\xi_{\mathbf{k}+\mathbf{Q}}}}%
{2})^{2}}$ and $E^{\pi}_{2}({\bf k}) =\frac{{\xi_{\mathbf{k}}+\xi_{\mathbf{k}+\mathbf{Q}}}}{2}%
+\sqrt{4{\delta}^{2}+(\frac{{\xi_{\mathbf{k}}-\xi_{\mathbf{k}+\mathbf{Q}}}}%
{2})^{2}}$. $\xi _{\mathbf{k}}=\varepsilon _{\mathbf{k}}+\Sigma _{\mathbf{k}}-\mu $
with the band energy $\varepsilon _{\mathbf{k}}=-2t(\cos {k_{x}a}+\cos {k_{y}a})-2t_{z}\cos {k_{z}a}_{z}$. The Hartree-Fock self-energy is given by $\Sigma _{\mathbf{k}}=V(0)n_{0}-{\frac{1}{N_L}}\sum_{\mathbf{k}^{\prime }}V({\mathbf{k}-\mathbf{k}%
^{\prime }})n_{\mathbf{k}^{\prime }}$ and $\delta\equiv\delta _{{\bf Q}}=\delta _{{-\bf Q}}$ describes the CDW order.

$\mathscr{H}^{\pi}_{BdG}$ clearly shows that the topology of the system can be engineered by simultaneously manipulating the two effective Hamiltonians $H'^{\pi}_{p-wave}$ and $H''^{\pi}_{p-wave}$, describing in-plane $p+ip$ superfluids.
This can be naturally achieved via  the spontaneously formed density modulation in our proposed scheme. To show this, let us consider the Hamiltonian
$\mathscr{H}^{\pi}_{BdG}$ at a fixed $k_z$. The topologically distinct regions of $p+ip$ superfluids Hamiltonian $H'^{\pi}_{p-wave}$ ($H''^{\pi}_{p-wave}$) are (i) $\mu'(\mu'') <-4t'$ and $\mu'(\mu'') > 4t'$, which are the topological trivial region, (ii) $-4t'<\mu'(\mu'')<0$ and $0<\mu'(\mu'')<4t'$, which are the topological regions with opposite chirality. Here $\mu'=\widetilde{\mu}{-}\sqrt{4{\delta}^{2}+
(-2t_{z}^{\prime}\cos{k_{z}a}_{z})^{2}}$ and $\mu''=\tilde{\mu}{+}\sqrt{4{\delta
}^{2}+(-2t_{z}^{\prime}\cos{k_{z}a}_{z})^{2}}$ with the effective hoppings $t_{\alpha}^{\prime}=t_{\alpha}-\Sigma_{\alpha}/2$,
$t^{\prime}\equiv t_{x}^{\prime}= t_{y}^{\prime}$ and $\tilde{\mu}=\mu-V(0)n_{0}$.
Note that to simplify the discussion we consider the strongest exchange interaction
energy between nearest neighbors as  $\Sigma_{x(y)}=\sum\limits_{\mathbf{k}}\frac{2J}{N_{L}}\cos(k_{x(y)}a)n_{\mathbf{k}}$
and $\Sigma_{z}=-\sum\limits_{\mathbf{k}}\frac{4Ja^3}{N_{L}a_{z}^3}\cos(k_{z}a_{z})n_{\mathbf{k}}$,
with $J\equiv|d^{\prime 2}/(ta^{3})|$ capturing the strength of dipolar interaction. The two effective chemical potentials $\mu'$ and $\mu''$ can be tuned simultaneously by changing the CDW order $\delta$ via varying $J$, and thus can manipulate the topology of the system. For example, considering the case of $J=0.8$ in Fig.~\ref{fig:phase}(a), the CDW order $\delta$ simultaneously tunes the two effective chemical potentials in different regions: (i) $-\pi/2 \leq k_{z}a_z<-k_{I}^{c}\simeq 0.39\pi$ or
$k_{I}^{c}<k_{z}a_z<\pi/2$, where $0<\mu
^{\prime}(\mu'')<4t^{\prime}$, $H'^{\pi}_{p-wave}$ and $H''^{\pi}_{p-wave}$ are thus simultaneously engineered in the same topological region with Chern number $C=-1$, (ii) $-k_{I}^{c}\ <k_{z}a_z<k_{I}^{c}$, where $-4t^{\prime}<\mu^{\prime}<0$ and $0<\mu''<4t^{\prime}$, $H'^{\pi}_{p-wave}$ and $H''^{\pi}_{p-wave}$ are thus tuned in topological regions with opposite chirality characterized by $C=\pm 1$.
Therefore, a topological phase (phase I in Fig.~\ref{fig:phase}(a)) characterized by the existence of a topological phase transition between two topological regions with $C=-2$ and $C=0$ along the $k_z$-axis is achieved. While increasing $J$, for instance $J=1.2$, the CDW order increases and plays a dominant role in tuning effective chemical potentials. Distinct from smaller $J$, here for each $k_z$ within the Brillouin zone (BZ), the effective chemical potentials are set in the same region as $-4t^{\prime}<\mu^{\prime}<0$ and $\mu''>4t^{\prime}$  .
$H'^{\pi}_{p-wave}$ and $H''^{\pi}_{p-wave}$ are thus simultaneously engineered in a topological region with $C=1$ and a non-topological region with $C=0$, respectively.  Therefore, a new topological phase (phase IV in Fig.~\ref{fig:phase}(a)) characterized by a uniform Chern number ($C=1$) is obtained. Using the same approach, other two topological phases (phase II, III in Fig.~\ref{fig:phase}(a)) can be determined. We also find that there is a threshold of $J$, below which $\delta =0$, the $p$-wave superfluid is favored. We can map out the zero-temperature phase diagram as shown in Fig.~\ref{fig:phase}(a). Such an analysis is readily generalizable to the case  ${\mathbf Q }= (0, 0, 2\pi/3a_z)$. Another four distinct topological phases have been obtained (see details in SM), indicating that our scheme provides a systematical way of engineering the topological nature of the system.

\textbf{A toolbox for engineering elementary fermions}

We now show how to engineer three kinds of elementary fermions, i.e., Dirac, Weyl and Majorana fermions, through
manipulating the topology of the system.  As shown in Fig.~\ref{fig:phase}(a), in topological phases I and III, there is a topological phase transition occurring along the $k_z$-axis. For example, in phase III the phase boundary along the $k_z$-direction corresponds with the emergence of two gapless points in quasiparticle excitations at $(\pi, \pi,\pm k_{III}^c)$, as shown in Fig.~\ref{fig:Weyl}(a). It turns out that these two gapless points are Weyl nodes. Close to the Weyl nodes, the effective Hamiltonian takes the form of a $2 \times 2$ Hamiltonian describing chiral Weyl fermions, which can be expressed as
$\widetilde{H}_{eff}^{\pi}=-\Delta(k_{x}a-\pi)\sigma_{x}-\Delta(k_{y}a-\pi)\sigma_{y}-\frac{\partial
{E}^{\pi}_{1}}{\partial k_{z}}\mid_{\mathbf{k}=(\pi/a,\pi/a,k_{III}^{c}/a_{z})}%
(k_{z}a_{z}-k_{III}^{c})\sigma_{z} \equiv {\mathbf{d}}\cdot{\mathbf{\sigma}}$ (see SM for details). The quasiparticle energy dispersion is linear around Weyl points. As shown in Fig.~\ref{fig:Weyl}(b) and (c), the Weyl nodes are hedgehog-like topological defects of the vector field ${\mathbf{d}}$, which are the point source of Berry flux in momentum space, with a topological invariant $N_{C}=\mp 1$. $N_{C}$ is defined by $N_{C}=\frac{%
1}{24\pi ^{2}}\epsilon _{\mu \nu \gamma \chi }$tr$\oint\nolimits_{\bar{\Sigma} }d%
\mathbb{S}^{\chi }\bar{G}\frac{\partial \bar{G}^{-1}}{\partial k_{\mu }}\bar{G}\frac{\partial
\bar{G}^{-1}}{\partial k_{\nu }}\bar{G}\frac{\partial \bar{G}^{-1}}{\partial k_{\gamma }}$,
where $\bar{G}^{-1}$ is the inverse Green's function for the quasiparticle excitation, $\bar{\Sigma}$ is a 3D surface around the isolated gapless points
and $\mbox{tr}$ stands for the trace over the relevant particle-hole degrees of
freedom. Quasiparticle excitations near the gapless points realize the long-sought
low-temperature analogs of Weyl fermions originally proposed in particle physics. These Weyl nodes are separated from each other in momentum space. They cannot be hybridized, which makes them indestructible, as they can only disappear by mutual annihilation of pairs with opposite topological
charges, which is distinct from the spectral-gap protection in insulating topological phases.

Furthermore, as shown in Fig.~\ref{fig:phase}(a) when varying $J$, the system undergoes phase transitions between various topological phases. The phase boundaries correspond with the gap closing in quasiparticle excitations. Interestingly, we find that these gapless points develop low-temperature analogs of Dirac topological defect. For example, when considering the phase boundary between III and IV,  the effective Hamiltonian around the gapless point can be expressed as $\widetilde{{H'}}_{eff}^{\pi}=-\Delta(k_{x}a-\pi)\sigma_{x}-\Delta(k_{y}a-\pi)\sigma_{y} \equiv {\mathbf{h}}\cdot {\mathbf{\sigma}}$ (see details in SM).  As shown in Fig.~\ref{fig:Dirac}(b), the vector field ${\mathbf{h}}$ forms a vortex structure in the momentum space. At the vortex core, the length of the vector vanishes, indicating the gap closing in
quasiparticle excitations. Therefore, it forms a Dirac topological defect, which is  confirmed from the calculation of the winding number $W=\frac{1}{2\pi}\oint d\mathbf{k}\left[\frac{h_{x}}{\left\vert \mathbf{h}\right\vert}\overset{\rightharpoonup}{\nabla}\frac{h_{y}}{\left\vert \mathbf{h}\right\vert }%
-\frac{h_{y}}{\left\vert \mathbf{h}\right\vert }\overset{\rightharpoonup}{\nabla}%
\frac{h_{x}}{\left\vert \mathbf{h}\right\vert }\right]$ being equal to $-1$.

\begin{figure}[t]
\begin{center}
\includegraphics[width=8.7cm]{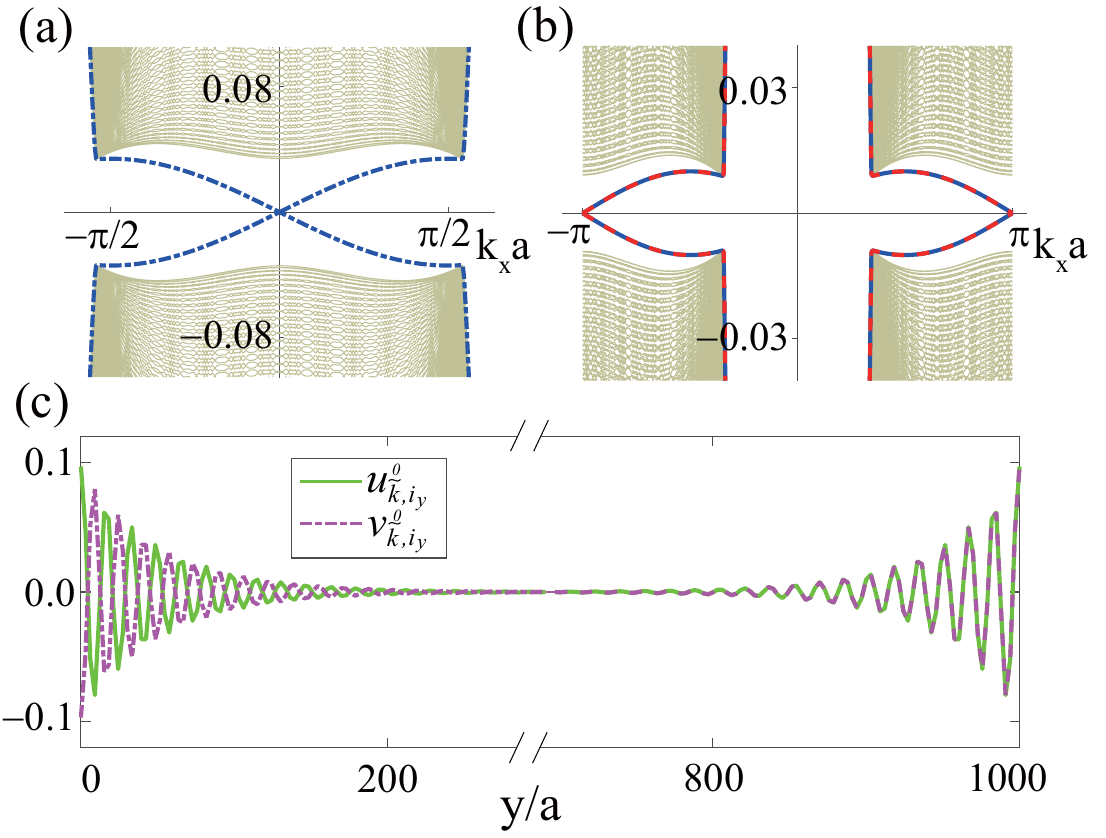}
\end{center}
\caption{(a)(b) Energy spectrum of the system with open (periodic) boundary conditions in the $y$ ($x$) directions for a fixed $k_z$. In (a), it is shown that there is one pair of chiral edge states marked by blue-dashed-line in phase IV with $J=2$, $n_0=0.6$ and $k_za_z=-\pi/2$. (b) shows that there are doubly degenerate chiral edge modes labeled by blue-dashed-line and red-solid-line in phase I with $J=0.93$, $n_0=0.66$ and $k_za_z=-\pi/2$. The wavefunction of the zero-energy state in (a) is shown in (c). It turns out that these zero-energy edge states are Majorana fermions. Other parameters are the same as in Fig.~\ref{fig:phase}.} \label{fig:MF}
\end{figure}

Besides hosting quasiparticles analogous to Dirac and Weyl fermions, our proposed set-up can also serve as a tool for systematically engineering both paired and unpaired Majorana fermions. For example, in phase IV there are two zero-energy states as shown in Fig.~\ref{fig:MF}(a) when $k_za_z=-\pi/2$. The corresponding wavefunctions (see details in SM) satisfy the relation $u_{\mathbf{\tilde{k}},i_{y}}^{0}=-v_{\mathbf{\widetilde{k}},i_{y}%
}^{0\ast}$ $[u_{\mathbf{\tilde{k}},i_{y}}^{0}=v_{\mathbf{\tilde{k}},i_{y}%
}^{0\ast}]$ on the left [right] edge (Fig.~\ref{fig:MF}(c)). These eigenstates thus support one localized Majorana fermion per edge of the system. More interestingly, the number of Majorana fermions shows high degree of tunability. Not only even number of Majorana fermions per edge, such as shown in Fig.~\ref{fig:MF}(b) (two per edge), but also odd number of Majorana fermions per edge, such as shown in Fig.~\ref{fig:MF} (a) (one per edge) and Fig. S6 (three per edge), can be hosted, which would supply new possibilities pointing to braiding statistics and applications to topological quantum computing.

\textbf{Estimates of the critical temperature}

In current experiments, for example, when considering $^{161}$Dy in a lattice with the lattice constant $a = 225nm$, where the dipolar interaction strength can be tuned as $J = 3$, the critical temperature of our proposed phases, such as shown in Fig.~\ref{fig:phase}(a), can reach around $0.1$nK (see SM for details). Furthermore, taking advantage of recent experimental realization of Feshbach resonance in magnetic lanthanide atoms~\cite{Baumann_PhysRevA,2014_Frisch_nature}, the dipole-dipole interaction becomes highly tunable. The critical temperature can be estimated to reach around $10$nK or even higher, making our proposal promising for experimental realization.

\textbf{Conclusion}

We have shown how to construct a toolbox for systematically engineering three distinct emergent elementary fermions by tuning the topology of the system. A new link between searching fundamental particles and topological phenomena, such as $p+ip$ superfluids, was built. It thus paves a new way for transforming the fermion's nature between various elementary fermions, with applications ranging from fundamental physics to quantum computing.


\end{document}